\documentclass[prl,unsortedaddress,superscriptaddress,showpacs,twocolumn,floatfix]{revtex4}

\usepackage{amsmath}
\usepackage{graphicx}

\begin{document}

\title{Measurement of $P_{\mu}\xi$ in Polarized Muon Decay}

\affiliation{University of Alberta, Edmonton, AB, T6G 2J1, Canada}
\affiliation{University of British Columbia, Vancouver, BC, V6T 1Z1, Canada}
\affiliation{Kurchatov Institute, Moscow, 123182, Russia}
\affiliation{University of Montreal, Montreal, QC, H3C 3J7, Canada}
\affiliation{University of Regina, Regina, SK, S4S 0A2, Canada}
\affiliation{Texas A\&M University, College Station, TX 77843, U.S.A.}
\affiliation{TRIUMF, Vancouver, BC, V6T 2A3, Canada}
\affiliation{Valparaiso University, Valparaiso, IN 46383, U.S.A.}

\author{B.~Jamieson}
\affiliation{University of British Columbia, Vancouver, BC, V6T 1Z1, Canada}

\author{R.~Bayes}
\altaffiliation[Affiliated with: ]{Univ.~of Victoria,
Victoria, BC.}
\affiliation{TRIUMF, Vancouver, BC, V6T 2A3, Canada}

\author{Yu.I.~Davydov}
\altaffiliation[Affiliated with: ]{Kurchatov Institute,
Moscow, Russia.}
\affiliation{TRIUMF, Vancouver, BC, V6T 2A3, Canada}

\author{P.~Depommier}
\affiliation{University of Montreal, Montreal, QC, H3C 3J7, Canada}

\author{J.~Doornbos}
\affiliation{TRIUMF, Vancouver, BC, V6T 2A3, Canada}

\author{W.~Faszer}
\affiliation{TRIUMF, Vancouver, BC, V6T 2A3, Canada}

\author{M.C.~Fujiwara}
\affiliation{TRIUMF, Vancouver, BC, V6T 2A3, Canada}

\author{C.A.~Gagliardi}
\affiliation{Texas A\&M University, College Station, TX 77843, U.S.A.}

\author{A.~Gaponenko}
\altaffiliation[Present address: ]{LBNL,
Berkeley, CA.}
\affiliation{University of Alberta, Edmonton, AB, T6G 2J1, Canada}

\author{D.R.~Gill}
\affiliation{TRIUMF, Vancouver, BC, V6T 2A3, Canada}

\author{P.~Gumplinger}
\affiliation{TRIUMF, Vancouver, BC, V6T 2A3, Canada}

\author{M.D.~Hasinoff}
\affiliation{University of British Columbia, Vancouver, BC, V6T 1Z1, Canada}

\author{R.S.~Henderson}
\affiliation{TRIUMF, Vancouver, BC, V6T 2A3, Canada}

\author{J.~Hu}
\affiliation{TRIUMF, Vancouver, BC, V6T 2A3, Canada}

\author{P.~Kitching}
\affiliation{TRIUMF, Vancouver, BC, V6T 2A3, Canada}

\author{D.D.~Koetke}
\affiliation{Valparaiso University, Valparaiso, IN 46383, U.S.A.}

\author{J.A.~Macdonald }
\altaffiliation[Deceased.]{} 
\affiliation{TRIUMF, Vancouver, BC, V6T 2A3, Canada}

\author{R.P.~MacDonald}
\affiliation{University of Alberta, Edmonton, AB, T6G 2J1, Canada}

\author{G.M.~Marshall}
\affiliation{TRIUMF, Vancouver, BC, V6T 2A3, Canada}

\author{E.L.~Mathie}
\affiliation{University of Regina, Regina, SK, S4S 0A2, Canada}

\author{R.E.~Mischke}
\affiliation{TRIUMF, Vancouver, BC, V6T 2A3, Canada}

\author{J.R.~Musser}
\altaffiliation[Present address: ]{Arkansas Tech University,
Russellville, AR.}
\affiliation{Texas A\&M University, College Station, TX 77843, U.S.A.}

\author{M.~Nozar}
\affiliation{TRIUMF, Vancouver, BC, V6T 2A3, Canada}

\author{K.~Olchanski}
\affiliation{TRIUMF, Vancouver, BC, V6T 2A3, Canada}

\author{A.~Olin}
\altaffiliation[Affiliated with: ]{Univ.~of Victoria,
Victoria, BC.}
\affiliation{TRIUMF, Vancouver, BC, V6T 2A3, Canada}

\author{R.~Openshaw}
\affiliation{TRIUMF, Vancouver, BC, V6T 2A3, Canada}

\author{T.A.~Porcelli}
\altaffiliation[Present address: ]{Univ.~of Manitoba,
Winnipeg, MB.}
\affiliation{TRIUMF, Vancouver, BC, V6T 2A3, Canada}

\author{J.-M.~Poutissou}
\affiliation{TRIUMF, Vancouver, BC, V6T 2A3, Canada}

\author{R.~Poutissou}
\affiliation{TRIUMF, Vancouver, BC, V6T 2A3, Canada}

\author{M.A.~Quraan}
\altaffiliation[Present address: ]{VSM Medtech Ltd., Coquitlam, BC.}
\affiliation{University of Alberta, Edmonton, AB, T6G 2J1, Canada}

\author{N.L.~Rodning}
\altaffiliation[Deceased.]{} 
\affiliation{University of Alberta, Edmonton, AB, T6G 2J1, Canada}

\author{V.~Selivanov}
\affiliation{Kurchatov Institute, Moscow, 123182, Russia}

\author{G.~Sheffer}
\affiliation{TRIUMF, Vancouver, BC, V6T 2A3, Canada}

\author{B.~Shin}
\altaffiliation[Affiliated with: ]{Univ.~of Saskatchewan,
Saskatoon, SK.}
\affiliation{TRIUMF, Vancouver, BC, V6T 2A3, Canada}

\author{T.D.S.~Stanislaus}
\affiliation{Valparaiso University, Valparaiso, IN 46383, U.S.A.}

\author{R.~Tacik}
\affiliation{University of Regina, Regina, SK, S4S 0A2, Canada}

\author{V.D.~Torokhov}
\affiliation{Kurchatov Institute, Moscow, 123182, Russia}

\author{R.E.~Tribble}
\affiliation{Texas A\&M University, College Station, TX 77843, U.S.A.}

\author{M.A.~Vasiliev}
\affiliation{Texas A\&M University, College Station, TX 77843, U.S.A.}

\collaboration{TWIST Collaboration}
\noaffiliation

\date{\today}

\begin{abstract}
  The quantity $P_{\mu}^{\pi}\xi$, where $\xi$ is one of the muon
  decay parameters and $P_{\mu}^{\pi}$ is the degree of muon
  polarization in pion decay, has been measured.  The value
  $P_{\mu}^{\pi}\xi=1.0003\pm0.0006\ ({\rm stat.})\pm0.0038\ ({\rm syst.})$ 
  was obtained.  This result agrees with previous
  measurements but is over a factor of two more precise.  It also
  agrees with the Standard Model prediction for $P_{\mu}^{\pi}\xi$ and
  thus leads to restrictions on left-right symmetric models.
\end{abstract}


\maketitle

\section{I. Introduction}

In the Standard Model (SM) of particle physics, positive muons decay via the weak
($V\!-\!A$) interaction into positrons plus neutrinos:
$\mu \! \rightarrow \! e\nu \overline{\nu}$ through a virtual state
involving $W$ vector bosons.  More generally, the amplitude for muon
decay can be described in terms of a local decay matrix element, which
is invariant under Lorentz transformations:

\begin{equation}
\label{eq:ldm}
M = \frac{4 G_F}{\sqrt{2}} \sum_{\substack{\gamma=S,V,T\\ \epsilon,\mu=R,L}} 
g^{\gamma}_{\epsilon\mu}
\langle \bar{e}_{\epsilon} | \Gamma^{\gamma} | \nu \rangle
\langle \overline{\nu} | \Gamma_{\gamma} | \mu_{\mu} \rangle,
\end{equation}

\noindent
where the $g^{\gamma}_{\epsilon\mu}$ specify the scalar, vector, and
tensor couplings between $\mu$-handed muons and $\epsilon$-handed
positrons \cite{Fetscher}.  In the SM $g^V_{LL}$ = 1, and all other
coupling constants are zero.

The differential decay spectrum \cite{Michel} of the $e^{+}$ emitted
in the decay of a polarized $\mu^{+}$ can be described by four
parameters -- $\rho$, $\delta$, $\eta$ and $\xi$ -- commonly referred
to as the Michel parameters, which are bilinear combinations of the
coupling constants.  In the limit where the positron and neutrino
masses are neglected, and radiative corrections \cite{Arbuzov} are not
explicitly included, this spectrum is given by:
\begin{center}
\begin{eqnarray}
\label{eq:decayrate}
    \frac{d^2 \Gamma}{dx d(\cos \theta)} \propto 3 (x^2-x^3)  +  \frac{2}{3} \rho (4x^3-3x^2) \nonumber \\ 
 +  P_{\mu} \xi \cos{\theta} ( x^2 - x^3 ) + P_{\mu} \xi \delta \cos{\theta} \frac{2}{3} (4x^3-3x^2),
\end{eqnarray}
\end{center}
where $\theta$ is the angle between the muon polarization and the
outgoing positron direction, $x=E_e/E_{max}$,
$E_{max}=(m_{\mu}^2+m_{e}^2)/2m_{\mu}=52.83$ MeV, and $P_{\mu}$ is the
degree of muon polarization.  The fourth parameter, $\eta$, appears in
the isotropic term when the positron mass is included in the analysis.
In the SM, the Michel parameters take on the precise values
$\rho=\delta=0.75$, $\xi=1$, and $\eta=0$.  The parameter $\xi$
expresses the level of parity violation in muon decay, while $\delta$
parametrizes its momentum dependence.

In this experiment $P_{\mu}$ is the magnitude of the $\mu^+$
polarization along the beam axis at the time of muon decay.  Surface
$\mu^+$ \cite{Pifer}, which are muons produced from $\pi^+$ decays at
rest, have a polarization of magnitude $P_{\mu}^{\pi}$, with a
direction antiparallel to their momentum, given by a generalization of
Eq. \ref{eq:ldm} for semileptonic decays.  In the SM with massless
neutrinos $P_{\mu}^{\pi}=1$.  In this experiment $P_{\mu}\xi$ is
determined from the positron spectrum, and then $P_{\mu}^{\pi}\xi$ is
obtained using the measurements of the muon trajectories.

SM extensions involving right-handed interactions \cite{Herczeg}
require deviations from pure $V\!-\!A$ coupling that can alter
$P_{\mu}\xi$.  Four probabilities
$Q_{\epsilon \mu}(\epsilon,\mu=R,L)$ for the decay of a $\mu$-handed
muon into an $\epsilon$-handed positron are given by:
\begin{equation}
Q_{\epsilon \mu} = \frac{1}{4} |g_{\epsilon\mu}^{S}|^{2} +
 |g_{\epsilon\mu}^{V}|^{2} +
3(1-\delta_{\epsilon\mu}) |g_{\epsilon\mu}^{T}|^{2},
\end{equation}
where $\delta_{\epsilon\mu}=1$ for $\epsilon=\mu$ and
$\delta_{\epsilon\mu}=0$ for $\epsilon\neq\mu$.  The probability:
\begin{eqnarray}
Q_{\epsilon R} & = & \frac{1}{4} |g^S_{LR}|^2 + \frac{1}{4} |g^S_{RR}|^2
+ |g^V_{LR}|^2 + |g^V_{RR}|^2 + 3 |g^T_{LR}|^2 \nonumber \\
 & = & \frac{1}{2} [ 1 + \frac{1}{3} \xi - \frac{16}{9} \xi\delta ],
\label{eq:RHcurrents}
\end{eqnarray}
sets a model independent limit on any muon right-handed couplings
\cite{Fetscher,PDG}. A recent review of muon decay is presented in
\cite{Kuno}.

A precision measurement of muon decay can place limits on left-right
symmetric (LRS) models \cite{Herczeg}.  The LRS models contain four
charged gauge bosons ($W^{\pm}_1$, $W^{\pm}_2$), the photon, and two
additional massive neutral gauge bosons.  The $W_1$ and $W_2$ masses
are $m_1$ and $m_2$ respectively, and the fields $W_L$ and $W_R$ are
related to the mass eigenstates $W_1$ and $W_2$ through a mixing angle
$\zeta$.  In these models both $V\!-\!A$ and $V\!+\!A$ couplings are
present, and parity violation appears because of the difference in the
mass of the vector bosons. In the SM the weak interaction contains
only two left handed vector bosons, $W_L^{\pm}$.

In the general LRS model \cite{Herczeg},
\begin{equation}
\xi \approx 1 - 2 \left[ \left( \frac{g_R}{g_L} \frac{m_1}{m_2} \right)^4 + \left(\frac{g_R}{g_L}\zeta \right)^2 \right] ,
\label{Eq:gen_LRS}
\end{equation}
where $g_R$ and $g_L$ are the right- and left-handed gauge couplings.
The manifest left-right symmetric model makes the additional
assumptions that $g_R = g_L$ and that the left- and right-handed quark
mixing matrices are identical.  In this case, $P_{\mu}^{\pi}$ can also be
expressed in terms of $m_1/m_2$ and $\zeta$, and one obtains
\cite{Herczeg}:
\begin{equation}
P_{\mu}^{\pi} \xi \approx 1 - 4 \left( \frac{m_1}{m_2} \right)^4 - 4 \zeta^2 - 4 \left( \frac{m_1}{m_2}\right)^2 \zeta.
\end{equation}

Prior to TWIST, the most precise direct measurement of $P_{\mu}^{\pi}\xi$
was $1.0027$ $\pm$ $0.0079$ (stat.) $\pm0.0028$ (syst.) \cite{Beltrami}, 
in agreement with the SM.  A similar value has been
measured using muons from kaon decay \cite{Imazato}.  Recently the
TWIST collaboration reported new measurements of 
$\rho = 0.75080$ $\pm$ $0.00032$ (stat.) $\pm$ $0.00097$ (syst.) $\pm$ $0.000023$ ($\eta$) \cite{Musser} 
and $\delta = 0.74964$ $\pm$ $0.00066$ (stat.) $\pm$ $0.00112$ (syst.) \cite{Gaponenko}.  
Using the result $P_{\mu}^{\pi}\xi\delta/\rho > 0.99682$, at the $90\%$
confidence level \cite{Jodidio}, along with the TWIST measurements of
$\rho$ and $\delta$, an indirect limit on $P_{\mu}^{\pi}\xi$ was determined
to be $0.9960<P_{\mu}^{\pi}\xi\leq\xi<1.0040$ ($90\%$ confidence level) \cite{Gaponenko}.
In this paper a new measurement of $P_{\mu}^{\pi}\xi$ is reported.

\section{II. Experimental Procedures}

In the present experiment, highly polarized surface muons \cite{Pifer}
were delivered, in vacuum, to the TWIST spectrometer \cite{Henderson}
by the M13 channel at TRIUMF \cite{Oram}.  The properties of the
surface muon beam are a typical rate of 2.5 kHz, a momentum of 29.6
MeV/c, and a momentum bite, $\Delta p/p\approx1.0\%$ FWHM.  Beam
positrons with a typical rate of 22 kHz, and the same momentum as the
muons are also a feature of the surface muon beam.

\begin{figure}[htbp]
 \includegraphics[width=1.0\columnwidth]{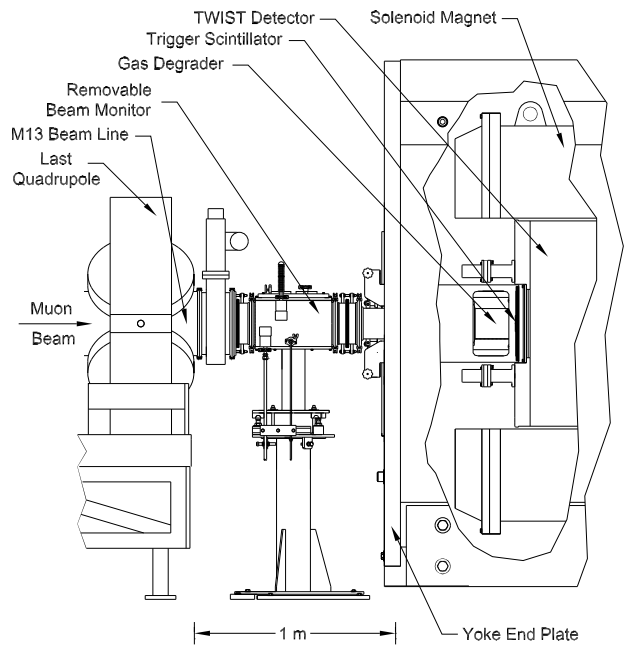}
\caption[Beam monitor]{\label{fig:beampackage} Location of the last
beamline quadrupole, beam monitor, gas degrader, trigger scintillator,
and the TWIST solenoid.  }
\end{figure}

A low pressure (8 kPa dimethyl ether gas) removable beam monitoring
chamber \cite{Sheffer} provides information on the muon beam before it
traverses the fringe field of the solenoid.  The chamber consists of
two 4 cm long modules, one to measure the position and direction of
the muon beam in the horizontal ($x$) direction, and the other for
measurements in the vertical ($y$) direction.  The location of the
final quadrupole of the M13 channel, the beam monitor, muon ranging
gas degrader, trigger scintillator and the TWIST detector are shown in
Fig. \ref{fig:beampackage}.  The beam monitor is inserted for
measurement of the beam properties, and removed during data
collection.

\begin{figure}[htbp]
\includegraphics[width=1.0\columnwidth]{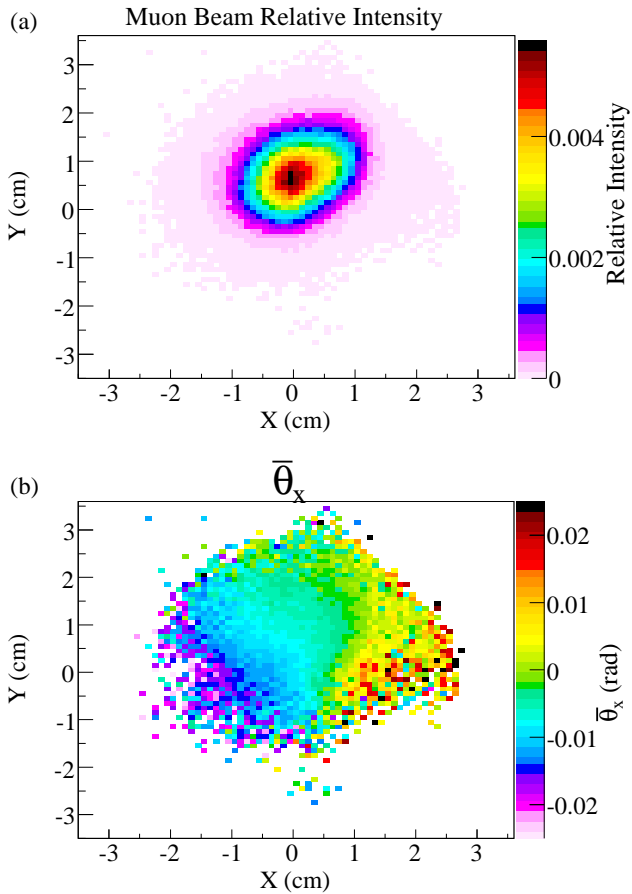}
\caption[Beam measurement]{\label{fig:beammeasurement} (colour online)
  Beam measurements projected to the center of the beam monitor at
  each 1 mm by 1 mm bin in $y$ versus $x$. (a) Muon beam intensity
  profile. (b) Mean angle in $x$ ($\bar{\theta_{x}}$).}
\end{figure}

\begin{figure}[htbp]
\includegraphics[width=1.0\columnwidth]{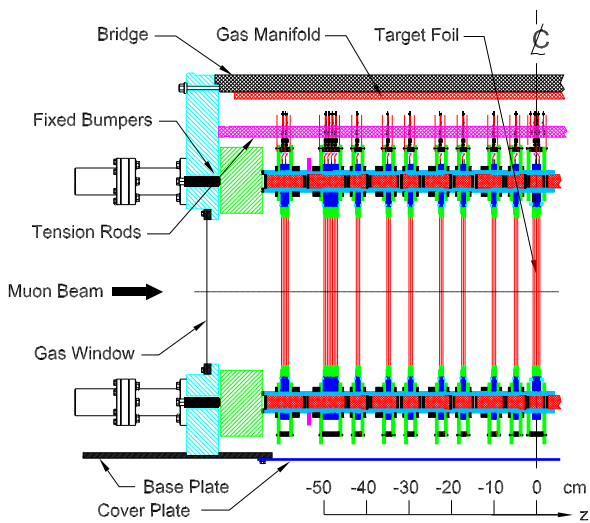}
\caption[Beam measurement]{\label{fig:stack} (colour online) Side view of the
upstream half of the TWIST spectrometer planar chambers and support
structure. }
\end{figure}

Changing the angle of the muon beam relative to the magnetic field
axis gives rise to a change in the polarization.  The solenoid field
is found to interact with the iron of the beamline magnets, such that
the muon beam is deflected off axis.  The available M13 channel
magnets could only partially alleviate this deflection.  Figure
\ref{fig:beammeasurement}(a) shows that the beam is centered in $x$,
but not in $y$.  The measured muon beam distribution in position and
angle at the beam monitor, as shown in Fig. \ref{fig:beammeasurement},
is input into a simulation to calculate the average depolarization of
the muons from the location of the beam monitor to the high-purity Al
muon stopping target at the center of the TWIST detector
(Fig. \ref{fig:stack}).  The angular distribution in $x$ is shown in
the figure, but similar distributions are measured in $y$ and these
are also included in the simulation \cite{Sheffer}.  The RMS size of the beam is 0.6
cm in both $x$ and $y$, and the average angular spread of the beam is
10 mrad in $x$ and 12 mrad in $y$.

The muon beam is transported in vacuum until it is inside the solenoid
field, where it then passes through the gas degrader and trigger
scintillator.  The gas degrader is a 21.67 cm long (along $z$) chamber
containing a mixture of He and CO$_2$ that can be adjusted to stop the
muons in the stopping target.  The plastic (Bicron BC408) disk shaped
trigger scintillator has a thickness of 195 $\mu$m, a radius of 3 cm,
and is located 80 cm upstream of the central stopping target.

The materials that the muons pass through, starting from the production target
to the center of the stopping target, are summarized in Table
\ref{tab:matbudget}.

\begin{table}[!hbt]
  \begin{center}
    \begin{tabular}{lr}
      \hline
      Material                             & Thickness (mg/cm$^2$) \\
      \hline
      Degrader and vacuum foils            & 11.91          \\
      He/CO$_2$ degrader                   &  1.95 to 42.80 \\
      Air gap ($2.82$ cm)                  &  3.65          \\
      Muon scintillator                    & 20.12          \\
      Scintillator wrap                    &  3.03          \\
      Cradle window                        &  0.88          \\
      PC module                            &  9.46          \\
      Dense stack                          & 13.41          \\
      Seven UV modules                     & 27.80          \\
      He/N$_2$ ($63.8$ cm)                 & 12.25          \\
      Half target module before target     &  4.31          \\
      Half of 71 $\mu$m Al target          &  9.59          \\
      \hline
      Total to center of Al target         & 118.36 to 159.21 \\
      \hline
    \end{tabular}
    \caption[Muon stopping materials budget]{ Estimates of the
      material thicknesses in mg/cm$^2$ that the muon penetrates from
      the production target to the center of the muon stopping target.
      Surface muon momenta are such that this approximately matches
      the muon range for a degrader containing $50$\% CO$_2$.}
    \label{tab:matbudget}
  \end{center}
\end{table}

The TWIST spectrometer \cite{Henderson} is designed to measure a broad
range of the positron spectrum from muon decays in a stopping target,
allowing the simultaneous extraction of the spectrum shape parameters.
The spectrometer consists of 12 very thin high-precision proportional
chamber (PC) planes and 44 drift chamber (DC) planes, perpendicular to
the axis of a solenoid producing a magnetic field of 2 T.  The
upstream half of the spectrometer (22 DCs and 6 PCs) is shown in
Fig. \ref{fig:stack}, and the downstream half of the detector is
mirror symmetric with the upstream.

The gas particle detectors used by TWIST are mounted in a leak tight
aluminum cradle that is filled with $3$\% N$_2$ and $97$\% He gas.
Helium, rather than air, is used to reduce the scattering between
measurement points and to allow the low energy muons to reach the
stopping target at the center of TWIST.  The spacing of the PCs and
DCs is maintained by precision ground spacers made of a Russian
engineered material Sital CO-113M which has a very small thermal
coefficient of expansion.  The spacers are pushed together by four
pneumatic cylinders with a force of $1470$ N each to ensure they are
touching and do not move.

The DCs are used to obtain a precise measurement of the position of
the e$^+$ as it traverses the chambers.  To accomplish this a slow
drift gas, dimethyl ether (DME), which has a small angle between the
drift field and the electron drift direction in a non-zero magnetic
field (Lorentz angle), is used.  Each DC plane consists of 80
gold-plated tungsten anode sense wires of $15$ $\mu$m diameter, spaced
by 4 mm \cite{Davydov}.  The cathode foil walls are $6.35$ $\mu$m of
aluminized Mylar, nominally placed $2$ mm from the wires.  The drift
time for each signal is used along with a space-to-time relationship
(STR) that is calculated by a GARFIELD simulation \cite{garfield} to
improve the determination of the momentum and angle of the decay
positrons.

The purpose of the PCs is to have reasonably fast timing information
to help with pattern recognition.  The width of the time to digital
converter (TDC) signal from the PCs is used to discriminate muons from
positrons.  The PCs are similar to the DCs, with the exception of the
following: the PCs use CF$_4$/isobutane gas, a wire spacing of 2mm,
and 160 wires in each plane.  The PC modules consist of four wire
planes.  One PC module is placed at either end of the detector stack,
and the specialized target module in the center of the detector stack
contains two PCs on either side of the muon stopping target.

Collection of an event is triggered by the muon scintillator whose
threshold is set high enough to reduce the beam positron trigger rate
but low enough to trigger on most beam muons.  The fraction of triggers
due to beam positrons is only $10$\%.  Positrons in the beam are
easily removed from the data since they leave signals through the full
length of the detector.

Events in TWIST are recorded by LeCroy Model 1877 TDCs. The trigger
and TDC read-out are set up to record signals in 500 ps time bins from
$6$ $\mu$s before to $10$ $\mu$s after a muon passes through the
trigger scintillator.  For each wire signal, the TDC records a pulse
start and stop, and thus can be used to calculate a pulse width.  In
this configuration up to eight wire signals can be recorded for each
wire in any triggered event \cite{daq-renee}.  A fixed blanking time
of 80 $\mu$s was imposed after each accepted trigger. This blank is
sufficiently long to allow each TDC to finish conversion. To insure
that all TDCs have fully converted their hits, each one was required
to produce a header specifying how many hits were converted even when
zero hits were received.

For this measurement a $71\pm1$ $\mu$m thick, $99.999$\% pure, Al
target is used to stop muons.  The tracks of the selected muons are
required to have at least one signal in the PC immediately before the
Al target and no signals in the PC immediately after the Al target.
The average $z$ from the last chamber plane fired by the muons
($\bar{z}_{\mu}$) is used in a feedback loop to control the fractions
of He and CO$_2$ in a gas degrader, in order to stop $97.0\pm1.5$\% of
the selected muons in the Al target at the center of the solenoid.
Some muons that stop in the CF$_4$/isobutane or Mylar foils in the
vicinity of the Al target could not be separated and wer included in
the measurement.

\section{III. Data Analysis}
TWIST determines the Michel parameters by fitting two-dimensional
distributions of reconstructed experimental decay positron momenta and
angles with distributions of reconstructed simulated data
\cite{Gaponenko}. 

The nominal fiducial region adopted for this analysis requires $p<50$
MeV/c, $|p_z|>13.7$ MeV/c, $p_T<38.5$ MeV/c, and
$0.50<|\cos\theta|<0.84$.  The fiducial cuts, while intentionally
chosen to be conservative, are related to physical limitations of the
TWIST detector. The 50 MeV/c momentum cut rejects events that are near
the region utilized in the energy calibration.  It is also important
to avoid the region very close to the end point to minimize the
sensitivity of the Michel parameter fits to details of the simulation
that may affect the momentum resolution.  The longitudinal momentum
constraint eliminates events with helix pitch near the 12.4 cm
periodicity in the wire chamber spacing.  The transverse momentum
constraint ensures that all decays are well confined within the wire
chamber volume.  The angular constraint removes events at large
$|\cos\theta|$ that have worse resolution and events at small
$|\cos\theta|$ that experience large energy loss and multiple
scattering.  These limits were fixed early in the analysis.  The value
of $P_{\mu}^{\pi}\xi$ was found to change by less than 0.0001 when the
fiducial boundaries were moved by $\pm 2$\% in momentum cut values and
$\pm 10$\% in $|\cos\theta|$ cut values.

The decay positrons spiral through the chambers producing signals on the
wires, which are recorded by the TDCs.  These helical tracks are
subsequently reconstructed and analyzed to determine the positron
energy and angular distributions.  Determining the positron momentum
and angle is done with a $\chi$-squared fit to a helical track that
includes the drift time information from each cell.  The helix fits
also include positron multiple scattering in the $\chi^2$ calculation
using the procedure from \cite{Lutz}.  The efficiency of the track
fitting is $\gtrsim$99.5\% within the nominal fiducial region used for
spectrum fitting \cite{Musser,Gaponenko}.  The number of degrees of
freedom (DOF) varies with track angle, since higher angle tracks will
pass through more DC cells per plane.  For the average DOF, the most
probable $\chi^2/DOF$ is $29.7/28 = 1.06$, but there is a long tail to
larger values because the non-Gaussian tails of the multiple
scattering distributions are not treated properly in the $\chi^2$
calculation and because we only use approximate STRs.  The momentum
resolution is typically 100~keV/c, and the $\cos\theta$ resolution is
about 0.005 \cite{Musser}.

In event selection, the PC and DC signals are examined to identify events
in which the muon stopped in the target, then decayed at least 1.05
$\mu$s, and no more than 9 $\mu$s, later.  The delay insures that the
PC and DC signals associated with the muon and decay positron do not
overlap.  Events are rejected if a second muon enters the
spectrometer, or if a beam positron passes through the spectrometer
within 1.05 $\mu$s of either the muon arrival or decay time.
Additional cuts include the muon flight time through the M13 beam line
and a requirement that the muon stopping location be within 2.5 cm of
the detector axis.  All events that pass these cuts are analyzed to
reconstruct the decay positron kinematics.

After track fitting, $\sim$ 2\% of the events contain additional tracks in
coincidence with the decay.  Extra tracks can arise from beam
particles that are not resolved in time, events that scatter within
the detector leading to two reconstructed track segments, and events
that include delta rays or decay positrons that backscatter from
material outside the detector volume.  Algorithms were developed to
identify events with backscatters and reject coincident beam particles
\cite{Gaponenko}.  

\begin{figure}[htbp]
\includegraphics[width=1.0\columnwidth]{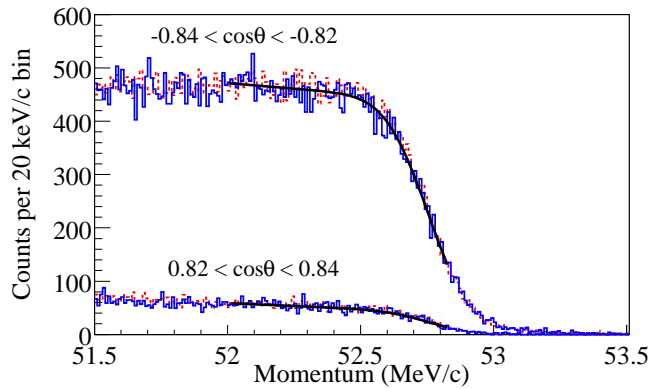}
\caption[End point fits]{\label{fig:ecfits}(colour online) Sections of
  the 2-dimensional end point fit function of the muon decay spectrum
  for the bins within the fiducial region containing the smallest
  upstream and downstream angles.  The data are shown as a solid-line,
  the matching simulation set is shown as a dashed-line, and the
  endpoint fit functions are shown as smooth curves.}
\label{fig:ecbins}
\end{figure}

The energy calibration of the decay positrons is obtained from a fit
to the endpoint of the spectrum.  The fit for the endpoint is done
separately from the fit for the muon decay parameters, and to avoid
bias the data from the region of momentum used in the endpoint fit are
excluded from the determination of the muon decay parameters. The
endpoint fit function is a slope with an edge ($f(x)=ax+b\ {\rm for}\
x\leq1,\ {\rm and}\ f(x)=0\ {\rm for}\ x>1$) convoluted with a
Gaussian $\sigma$. The parameters $a$, $b$, and $\sigma$ all depend on
$\cos\theta$ \cite{andrthesis}.  
The end point of the muon decay spectrum and sections of the
2-dimensional end point fit function for the bins within the fiducial
region with the smallest upstream and downstream angles are shown in
Fig. \ref{fig:ecbins}.  The difference in yield between upstream and
downstream emphasizes the asymmetry of polarized muon decay.  The
corrected momentum $p_{\rm ec}$ is given by:
\begin{equation}
p_{\rm ec} = p_{\rm rec} \left( 1+\frac{\beta}{p_{\rm edge}} \right)^{-1} + 
\frac{\alpha}{|\cos{\theta}|},
\label{eqnecal}
\end{equation}
where $p_{\rm rec}$ is the reconstructed momentum, $\cos{\theta}$ is
the reconstructed cosine of the decay positron angle, $p_{\rm edge}$
is the maximum positron momentum, $\beta$ defines the momentum scale
related to the magnitude of the spectrometer magnetic field, and
$\alpha=(\alpha_u,\alpha_d)$ is the zero angle energy loss for
upstream ($u$) or downstream ($d$) decay positron tracks.  This simple
form was chosen because of the planar geometry of the wire chambers,
such that the amount of material the positron passes through increases
linearly as $1/\cos{\theta}$.

The endpoint fit parameter $\beta$ is highly correlated with
$\alpha_{\rm sum}=\alpha_u+\alpha_d$.  In the simulation the momentum
is measured without bias at the few keV/c level, the magnetic field is
measured with an NMR probe, and the magnetic field map has been
determined to better than $0.2$ mT in the tracking region.  For this
reason the endpoint fits are done with the value of $\beta$ set to
zero and assigned an uncertainty consistent with the momentum fit and
field map accuracy.  A 12 keV/c difference between the data and
simulation $\alpha_{\rm sum}$ values was observed, and this is
corrected by applying the energy calibration.

The TWIST simulation is based on GEANT 3.21 \cite{Brun}.  The
simulation contains virtually all the components of the spectrometer
with which a muon or a decay positron could interact. The output
exactly mimics the binary files generated by the data acquisition
system.  Details of the simulation have been presented previously
\cite{Musser,Gaponenko}.

In bringing the muon beam to a stop in the Al target at the center of
the TWIST spectrometer, the muons are depolarized by the combined
effect of multiple scattering, interaction with the fringe field of
the spectrometer, and interactions when stopped in the high purity Al
target.  Thus, the polarization of the muon with respect to the $z$
axis when it decays is lower than its polarization with respect to the
muon momentum when it was produced in pion decay.  To obtain an
absolute measurement of $P_{\mu}^{\pi}\xi$, data are fit to a
simulation that includes effects of fringe field depolarization and
material depolarization.

The main factors that influence the difference between $P_{\mu}^{\pi}$
and $P_{\mu}$ are the transport of the muon spins in the various
regions of magnetic field and models for the muon depolarization in
materials.  Transport of the muon spins from the beam measurement
location to the muon stopping target is done using a classical fourth
order Runge-Kutta using the Nystrom algorithm \cite{nystrom}, of the
Bargmann-Michel-Telegdi \cite{bargmann} equation.  The inherent
accuracy of this numerical integration is excellent, but depends on
the accuracy on the knowledge of the input beam parameters and the
magnetic field map.  For this reason, the beam monitoring chamber
measurements were used to generate the muon beam for the simulation.

Ideally the magnetic field is meant to be uniform and aligned with the
axis of the muon beam momentum (and anti-aligned with the muon spin),
which is labeled as the $+z$ coordinate in TWIST.  The field shaping
elements of the magnet were input into a finite element analysis which
models the magnetic field in order to create a field map that included
the radial components of the field as well as the longitudinal
components at each of the positions at which the longitudinal
component was measured.

The magnetic field of the solenoid was mapped using a specially
constructed mapping tool which used Hall probes, and NMR magnetometers
to measure the magnetic field.  The component of the magnetic field
along $z$ was mapped in the uniform field region and in the beam
monitoring entrance region.  A field map was generated that matches
the measurements to $0.2$ mT in $2.0$ T in the uniform region, but
deviates up to 4\% in the entrance region at $z<-200$ cm.  The effects
of the steel from the last two quadrupoles of the M13 beamline were
studied using field maps from the finite element analysis that
included or excluded the iron of the last three beamline elements.
The map produced from this study was used to estimate a systematic
error on $P_{\mu}^{\pi}\xi$ due to the uncertainty in the magnetic
field in the fringe field region.

A blind analysis was implemented by utilizing hidden Michel
parameters $\rho_H, \delta_H$, and $\xi_H$ to generate the simulated
decay rate.  The decay rate can be written as:
\[
\left. \frac{d^2 \Gamma}{dx d(\cos \theta)}\right|_{\rho_H, \delta_H, \xi_H}
+ \sum_{\lambda=\rho, \xi, \xi\delta} \frac{\partial}{\partial\lambda} \left[
  \frac{d^2 \Gamma} {dx d(\cos \theta)}\right] \Delta \lambda, 
\]
because the decay spectrum is linear in the shape parameters.  The
simulation spectrum was fit to the data spectrum by adjusting
$\Delta\rho$, $\Delta\xi$, and $\Delta\xi\delta$.

An alternate analysis scheme, used only to compare relative
polarizations, was developed using an integral asymmetry defined as
the difference between the number of forward and the number of
backward decays divided by their sum.  To obtain a polarization
estimate the forward and backward sums were done inside the fiducial
region described earlier in this section and normalized using
integrals of Eq. \ref{eq:decayrate} with the SM values of the
Michel parameters inserted.

\section{IV. Evaluation of Systematics}

The leading systematic uncertainties in this measurement of
$P_{\mu}^{\pi}\xi$ arise from the potential sources of muon depolarization.
These include depolarization due to multiple scattering in the
production target and a $3$ $\mu$m beamline foil, fringe field
depolarization, and interactions with material while the muon is
propagating through the detector and after stopping.

The depolarization in the production target is due to multiple
scattering of the muons while exiting the target.  An estimate of the
depolarization in the small angle approximation is $(\theta_{\rm
space}^{\rm RMS})^2/2$, where $\theta_{\rm space}^{\rm RMS}$ is the
width of the multiple scattering angular distribution \cite{PDG}.  The
muons in the beam arise from a maximum depth of $0.003$ cm of
graphite, which produces a depolarization of $0.2\times10^{-3}$ which
is taken as the systematic uncertainty in $P_{\mu}^{\pi}\xi$ due to
multiple scattering in the production target.

The systematic uncertainty in the fringe field depolarization is
estimated from the different settings used in data taking for the
second dipole element (B2) in the M13 channel.  The beam parameters
were measured for two different B2 settings, both before and after the
data collection.  The relative changes in angle and position between
the nominal B2 value (94.4 mT) and B2+0.5\% settings are similar for
the two periods, but the absolute numbers for the average beam angles
are quite different.  This could be due to changes in the performance
of the beam monitoring chamber or to its alignment to the beamline.
To determine the sensitivity of the polarization to beam position and
angle, a simulated beam was scanned in position and angle and the
polarization was found to depend quadratically on the input variables.
Using this parameterization, the predicted polarizations for the four
characterization runs are shown in Table \ref{tab:tecbeamaverages}.
The beam for the third row of Table \ref{tab:tecbeamaverages} is
illustrated in Fig. \ref{fig:beammeasurement}. The larger of the
differences in predicted polarization for a given B2 setting
($0.0033$) is adopted as an estimate of the uncertainty due to limits
of reproducibility.
\begin{table}[htb]
\caption{\label{tab:tecbeamaverages} Average beam positions and angles
  from beam monitoring measurements taken at different times, along
  with the simulation estimates of the muon polarization.  The first
  (last) two entries are from before (after) the data collection.}
\begin{ruledtabular}
\begin{tabular}{cccccc}
B2   &  $\bar{x}$ & $\bar{\theta_{x}}$ & $\bar{y}$ & $\bar{\theta_{y}}$  & $P_{\mu}^{\rm sim}$ \\ 
(mT) & (cm)     & (mrad)   & (cm)    & (mrad)    &           \\ \hline
94.4 &  0.07    &  -5.9    & 0.97    &   7.0     & 0.9929   \\
94.9 &  0.85    &  -1.1    & 0.87    &  -5.0     & 0.9955   \\
94.4 &  0.06    &  -6.7    & 0.73    & -11.2     & 0.9941   \\
94.9 &  0.94    &  -1.5    & 0.64    & -19.2     & 0.9922   \\
\end{tabular}
\end{ruledtabular}
\end{table}

Uncertainties due to deconvolution of the beam angle measurement,
modeling of the shape of the solenoid fringe field, and beam size
reproduction also contribute to the final quoted systematic
uncertainty of $0.0034$ due to fringe field depolarization.

The depolarization of the muons while they propagate through the
detector and interact with the detector materials is negligibly small
for muons with kinetic energy between $4$ MeV and $100$
keV \cite{senba}.  Simulations show that greater than $99$\% of muons
stopping in our Al target have more than $100$ keV kinetic
energy. Muonium formation is thus suppressed by ensuring that the
majority of the muons have sufficient energy entering the Al target.
For those muons that do not stop in the Al target, the Paschen-Back
effect minimizes but does not eliminate the depolarization due to
muonium formation.  Most of the muons stop in the high-purity Al
target, where they can interact with conduction electrons.  These
electrons create a hyperfine magnetic field at the site of the muon,
which can be considered as a fluctuating local field with a
correlation time $\tau_c~\simeq~10^{-13}$ s in Al \cite{mudepol:abragam}.  
This interaction results in a Korringa
depolarization rate \cite{mudepol:korringa,mudepol:cox} that has an
exponential form, and does not depend on the magnetic field.
Significant depolarization rates of $\lambda~>~0.001$ $\mu$s$^{-1}$
have been measured for muons in Cd, Sn, Pb, As, Sb, and Bi
\cite{mudepol:cox}.  The authors explained the measured depolarization
rates to be due to Korringa depolarization because the $\lambda$
values increase with temperature as predicted.

Jodidio \emph{et al.} \cite{Jodidio} measured a depolarization rate of
$(0.43\pm0.34)\times10^{-3}$ $\mu$s$^{-1}$ for their Al target at
$1.1$ T.  This rate is about 2.5$\sigma$ smaller than the
$(1.55\pm0.28)\times10^{-3}$ $\mu$s$^{-1}$ observed in this
experiment.  The difference could partly be due to the $2.5$ to
$5.5$\% of the muons that stop in the gas before the stopping target.
The functional form of the depolarization in gases is unknown, thus an
assumption that there is no unseen rapid depolarization is made. The
spin relaxation, with mean lattice-site residence time $\tau_c$, is
given approximately by the Kubo-Tomita expression \cite{kubo}, which
reduces to Gaussian (exponential) forms for $\tau_c \rightarrow \infty (\tau_c \rightarrow 0)$.

\begin{figure}[ht]
\center{\includegraphics*[width=1.0\columnwidth]{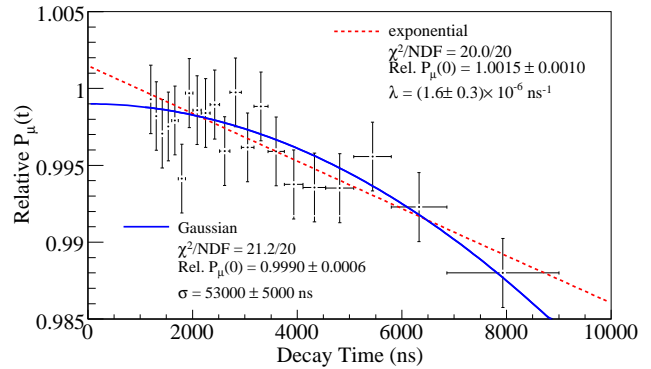}}
\caption{\label{fig:extrap} (colour online) Extrapolation to zero decay
  time of relative muon polarization, estimated using the decay
  positron integral asymmetry described in the text.  The
  extrapolation function is fit to data: with an exponential shown as
  a dashed-line, and as a Gaussian shown as a solid-line.}
\end{figure}

The difference between Gaussian and exponential extrapolations of the
integral asymmetry measurement, as shown in Fig. \ref{fig:extrap}, is
$2.4\times10^{-3}$.  Data before 1.05 $\mu$s are not considered because
of possible contamination of late TDC signals from muons for upstream
decay positrons.  Changing the cutoff between 900 ns and 1100 ns has
negligible effect on the asymmetry measurement extrapolated to zero
time.  Half the difference between the two different extrapolations is
the {\it correction} applied to the simulation to data fits, because
the simulation was generated with a Gaussian form, while in reality
the shape is most likely a linear combination of a Gaussian and
exponential.  An estimate of the extrapolation uncertainty is half the
difference between the Gaussian and exponential extrapolations.

Other systematic uncertainties were studied by employing the fitting
technique described in the data analysis section.  In this
case the fits are of experimental data (or simulation) samples, taken
with a systematic parameter set at an exaggerated level, to data (or
simulation) taken under ideal conditions.  The difference measured, or
sensitivity, expresses the changes in the spectrum shape caused by the
systematic effect in terms of the changes in the Michel parameters.
Systematic uncertainties in the measurement of $P_{\mu}^{\pi}\xi$ are
summarized in Table \ref{tbl:systematics}.

The chamber response systematic uncertainty in $P_{\mu}^{\pi}\xi$
comes from the uncertainty due to the PC and DC response.  Six sources
of uncertainty, described in the following paragraphs, are: time
variations in the wire-to-wire timing, changes to the space-to-time
relations (STR) from density variations, chamber foil bulge due to
changes in differential pressure between the chamber gas and cradle
helium, the asymmetry of the wire positions relative to the foils due
to construction, and a dead zone due to a recovery time after a muon
passes through the chambers.

To estimate the possible time variation in the wire to wire time zero
($t_0$), calibration runs are taken at the beginning and end of the
run period.  To measure the sensitivity of the muon decay parameters
to $t_0$ variations, a calibration scaled by ten times the difference
in the beginning and end of run time calibrations is used.  The
contribution to the chamber response systematic uncertainty in
$P_{\mu}^{\pi}\xi$ due to $t_0$ variations, obtained from fitting the
exaggerated time shift analysis spectrum to a nominal spectrum, is
$0.89 \times 10^{-3}$.  A second calibration scaled by 5 times the
difference in beginning and end of run, shows that this systematic
uncertainty scales linearly.

The shape of the chamber cathode foils depends on the differential
pressure between the chamber gas and the He gas of the cradle holding
all of the chambers.  The differential pressure of the chambers is
monitored and controlled by the gas system.  To measure the level of
foil bulge, several nominal data runs are taken at different
differential pressures.  The average helix fit drift time difference,
for tracks that go through the center of the chambers, relative to the
average helix fit drift times for tracks that go through the radially
distant part of the chamber, is sensitive to the foil bulge.  Run to
run monitoring of this fit time difference shows that the variation in
foil bulge is controlled to better than $50$ $\mu$m.  To measure the
sensitivity of the muon decay parameters to the chamber foil bulge, a
simulation generated with the STR calibration changed as if the
foils were moved outward by $500$ $\mu$m is used.  The contribution to
the chamber response systematic uncertainty, in $P_{\mu}^{\pi}\xi$,
due to time variations in the foil bulge, is estimated to be
$0.22\times10^{-3}$.

Surveys of the constructed chambers show that the spacing between the
wire plane and cathode foils is not the same on both sides of the
wire.  On average the cathodes are found to be shifted by about $150$
$\mu$m from their nominal 2 mm spacing from the wire planes.  With the
solenoid field on, the position of the cathode foils was also
determined using normal decay positron data.  Using the bulge
calibration data, from runs with different bulges, a relationship
between a fit time difference to a foil shift is obtained.  STR files
were changed half plane by half plane to match the measured shifts.  A
fit of the muon decay distribution, from the shifted cathode foil
simulation to a nominal simulation, results in a contribution to the
chamber response systematic uncertainty in $P_{\mu}^{\pi}\xi$ of
$0.2\times 10^{-3}$.

History plots of the density variation for each data set show that the
largest RMS change in density is $\pm 0.7$\%.  To estimate the effect
of density changes, a simulation was generated with the temperature
changed by $10\%$ (from 300 K to 270 K).  A scaled fit of the changed
temperature simulation set to a nominal simulation set is used to
estimate the addition to the chamber response systematic uncertainty
in $P_{\mu}^{\pi}\xi$ of $0.17\times 10^{-3}$.

As the muon passes through the upstream half of the detector, it
deposits large ionization in the chambers. The resulting electron
avalanches near the wires take time to recover as the positive ions
drift back to the cathode planes. These dead zones produce an
upstream-downstream efficiency difference for decay positrons. For
each muon hit the wire efficiency versus distance away from the muon
hit is estimated as a function of time after the muon track.  A
determination of the muon decay parameter sensitivities to the muon
dead zone is made by introducing an exaggerated dead zone into the
simulation which removes 11.4 times as many hits as are expected in
normal data.  A conservative estimate of the contribution to the
chamber response systematic uncertainty in $P_{\mu}^{\pi}\xi$ due to
the muon dead zone is $0.01\times10^{-3}$.

The systematic uncertainty in $P_{\mu}^{\pi}\xi$ due to spectrometer
alignment comes from how well the chamber translations in $x$, $y$,
$z$ and angle are corrected to match their true positions, and the
degree to which the misalignment between the magnetic field axis to
the chamber axis is treated.

The systematic uncertainty in $P_{\mu}^{\pi}\xi$ due to positron
interactions includes four effects: a discrepancy between the
simulation and data energy loss, hard and intermediate interactions,
multiple scattering, and backscattering from material outside the
detector.

The systematic uncertainty in $P_{\mu}^{\pi}\xi$ due to the momentum
calibration comes from two contributions.  One contribution is due to
how well the endpoint energy calibration can be determined.  The other factor
is how well the measured magnetic field map represents reality.

The systematic uncertainty in $P_{\mu}^{\pi}\xi$ due to a difference
in the upstream and downstream efficiencies is estimated from the mean
number of hits used in fitting the upstream and downstream tracks. An
upstream-downstream difference is expected from the different angular
distributions, and is reproduced by the simulation. The observed
discrepancy between data and simulation of 0.18 hits/track is a measure of
the efficiency difference.  

To estimate the sensitivity of the muon decay parameters, due to a
difference in upstream versus downstream efficiency, a data set was
analyzed with $5\%$ of its downstream DC hits thrown away before
analysis.  This upstream-downstream efficiency change, compared to a
normal analysis of the same data, produces a change in
$P_{\mu}^{\pi}\xi$ of ($1.9\pm0.9$)$\times 10^{-3}$.  The lowered
efficiency run downstream track fits have 1.8 fewer hits/track
relative to the standard run.  An estimate of the systematic
uncertainty in $P_{\mu}^{\pi}\xi$ due to a difference in upstream and
downstream efficiency is ergo $0.2\times10^{-3}$.

The systematic uncertainty in $P_{\mu}^{\pi}\xi$ due to background muon
contamination comes from muons which are born in pion decay in the
vicinity of the gas degrader.

The beam intensity systematic uncertainty in $P_{\mu}^{\pi}\xi$ comes from
changes in the beam rate which affects the rate of coincident particles.

Theoretical uncertainties in the radiative corrections introduce a
small systematic uncertainty in $P_{\mu}^{\pi}\xi$.

Several of the systematic uncertainties could vary from data set to
data set and are denoted by (ave), and are considered data set
dependent when calculating the weighted average value of $P_{\mu}^{\pi}\xi$.
For example, the effect of positron interactions on upstream and
downstream decay positrons changes when the mean muon stopping
location is adjusted; thus the systematic uncertainty in $P_{\mu}^{\pi}\xi$
due to positron interactions is set-dependent.

\begin{table}
\caption{\label{tbl:systematics} Contributions to the systematic
uncertainty for $P_{\mu}^{\pi}\xi$.  }
\begin{ruledtabular}
\begin{tabular}{lc}
 Effect                                   & Uncertainty \\ \hline
Depolarization in fringe field (ave)      & 0.0034 \\
Depolarization in stopping material (ave) & 0.0012 \\
Chamber response (ave)                    & 0.0010 \\
Spectrometer alignment                    & 0.0003 \\
Positron interactions(ave)                & 0.0003 \\
Depolarization in production target       & 0.0002 \\
Momentum calibration                      & 0.0002 \\
Upstream-downstream efficiency            & 0.0002 \\
Background muon contamination (ave)       & 0.0002 \\
Beam intensity (ave)                      & 0.0002 \\
Michel parameter $\eta$                   & 0.0001 \\
Theoretical radiative corrections         & 0.0001 \\
\end{tabular}
\end{ruledtabular}
\end{table}

\section{V. Results}
The result for $P_{\mu}^{\pi}\xi$ presented here uses a data sample
consisting of $2\times10^{9}$ events recorded in Fall 2004.  This data
sample includes eight data sets, of which seven were used for the
extraction of $P_{\mu}^{\pi}\xi$.  Simulations to fit each of the seven data
sets used different beam characterization profiles, derived from beam
measurements performed after the data collection, which matched
different conditions under which the data were recorded.  The
remaining data set was used to determine the detector response using
decay positrons from muons stopping in the trigger scintillator and
the first few chamber planes (far upstream), as described in
\cite{Musser,Gaponenko}.

Five sets of data were taken with the beam steered nominally.  One
data set had the muon beam stopping with the Bragg peak centered in
the target (stop $\frac{1}{2}$).  Two sets, which were separated in
time by a few days, were taken with the muon Bragg peak shifted to
$3/4$ of the way through the Al stopping target (stop $\frac{3}{4}$ A,
B).  One set was taken with a muon beam size limiting aperture
(aperture), and one set was taken with the beam rate increased (high
rate).  

Two sets of data were collected with the beam displaced by changing
the last bending magnet (B2) field by +0.5\% from nominal.  One of the
data sets (B2+0.5\%) had the muon Bragg peak centered in the stopping
target, while in the other set (PC5 stop), the muons were stopped
relatively far upstream in order to increase the relative fraction of
muons stopping in gas.  All of these data sets, using different beam
characterization profiles that matched the different conditions, were
used in this determination of $P_{\mu}^{\pi}\xi$.

The spectrum fit results for the parameter $P_{\mu}^{\pi}\xi$ are presented
in Table \ref{tbl:results}.  At the present stage TWIST cannot provide
an improved measurement of $\eta$, therefore its value is set to the
global analysis value of $-0.0036$ \cite{Gagliardi}, to constrain the
other parameters better.  The uncertainty of $\pm0.0069$ on the
accepted value of $\eta$ gives an uncertainty of $\pm0.0001$ on the
final value of $P_{\mu}^{\pi}\xi$.

\begin{table}
\caption{\label{tbl:results} Results for $P_{\mu}^{\pi}\xi$. Each fit has
1887 degrees of freedom. Statistical and set-dependent systematic
uncertainties are shown.  A description of the data sets is in the
text. }
\begin{ruledtabular}
\begin{tabular}{lcc}
Data Set                     & $P_{\mu}^{\pi} \xi$ $\pm$ stat $\pm$ syst & $\chi^2$       \\
\hline
B2+0.5\%                     & 1.0023 $\pm$ 0.0015 $\pm$ 0.0037  & 2007        \\
PC5 stop                     & 1.0055 $\pm$ 0.0030 $\pm$ 0.0038  & 1906        \\
stop $\frac{1}{2}$           & 1.0015 $\pm$ 0.0014 $\pm$ 0.0037  & 1876        \\
stop $\frac{3}{4}$ A         & 0.9961 $\pm$ 0.0014 $\pm$ 0.0037  & 1900        \\
high rate                    & 0.9997 $\pm$ 0.0019 $\pm$ 0.0037  & 1932        \\
aperture                     & 0.9978 $\pm$ 0.0018 $\pm$ 0.0037  & 1896        \\
stop $\frac{3}{4}$ B         & 1.0009 $\pm$ 0.0019 $\pm$ 0.0037  & 1841        \\
\end{tabular}
\end{ruledtabular}
\end{table}

The average values of $\rho$ and $\delta$ from the present fits are
0.749 and 0.753, respectively.  An evaluation of the uncertainties in
$\rho$ and $\delta$ has not been performed, but if one assumes
systematic uncertainties similar to the previous TWIST measurements,
these values are reasonably consistent with the published values of
$\rho = 0.7508$ \cite{Musser} and $\delta = 0.7496$ \cite{Gaponenko}.

\begin{figure}[ht]
\includegraphics*[width=1.0\columnwidth]{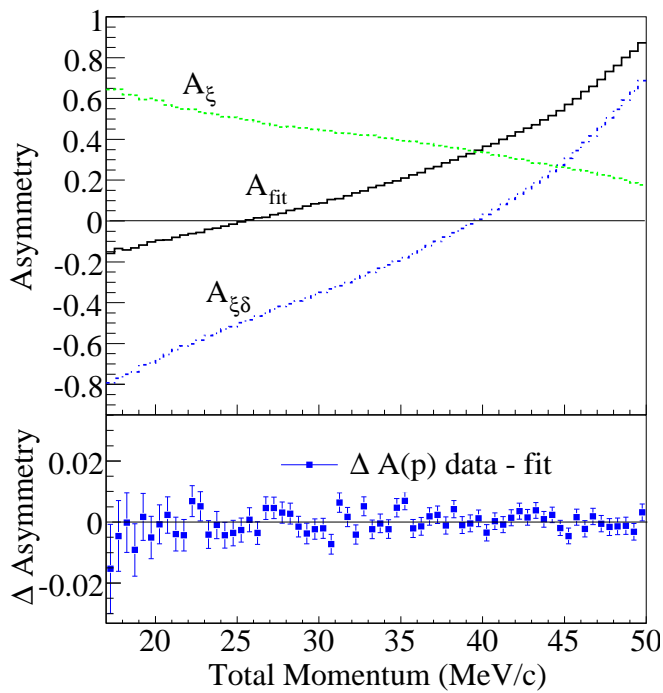}
\caption{\label{fig:ap} (colour online) The top panel shows the fit
 asymmetry versus positron momentum, $A(p)$, along with the
 contributions to the fit $A(p)$ from the $\xi$ and $\xi\delta$ terms.  The
 bottom panel shows the difference between the data and fit, $\Delta
 A(p)$.}
\end{figure}

To illustrate the quality of the fit, and how the spectrum fit
distinguishes between $P_{\mu}^{\pi}\xi$ and $P_{\mu}^{\pi}\xi\delta$, the
contribution to the fit asymmetry versus momentum for each of these
terms and from the best fit $A(p)$ are shown in Fig. \ref{fig:ap}.
Note that the total asymmetry versus momentum, $A(p)$, is:
\begin{equation}
A(p) = A_{\xi}(p) + A_{\xi\delta}(p).
\end{equation}
where $A_{\xi}$ is the asymmetry when the $P_{\mu}^{\pi} \xi \delta \cos{\theta} \frac{2}{3} (4x^3-3x^2)$ 
contribution to the positron
decay spectrum is zero, and $A_{\xi\delta}$ is the asymmetry when the
$P_{\mu}^{\pi} \xi \cos{\theta} ( x^2 - x^3 )$ contribution is zero.  The
top panel in Fig. \ref{fig:ap} shows the best fit asymmetry versus
positron momentum, $A(p)$ with all of the fiducial cuts applied as a
solid line; the contribution to the fit from the $\xi$ term as a
long-dashed line; and the contribution to the fit from the $\xi\delta$
term as the short-dashed line.  The bottom panel shows the difference,
$\Delta A(p)$ between data and fit.

\section{VI. Conclusion}

The value of $P_{\mu}^{\pi} \xi$ was determined to be 1.0003 $\pm$ 0.0006(stat.)
$\pm$ 0.0038(syst.).  The central value for $P_{\mu}^{\pi} \xi$ was
calculated as a weighted average using a quadratic sum of the
statistical and set-dependent uncertainties for the weights.  The
final systematic uncertainty is a quadratic sum of the set-independent
and the average values of the set-dependent systematics.

\begin{figure}[!htb]
  \includegraphics[width=1.0\columnwidth]{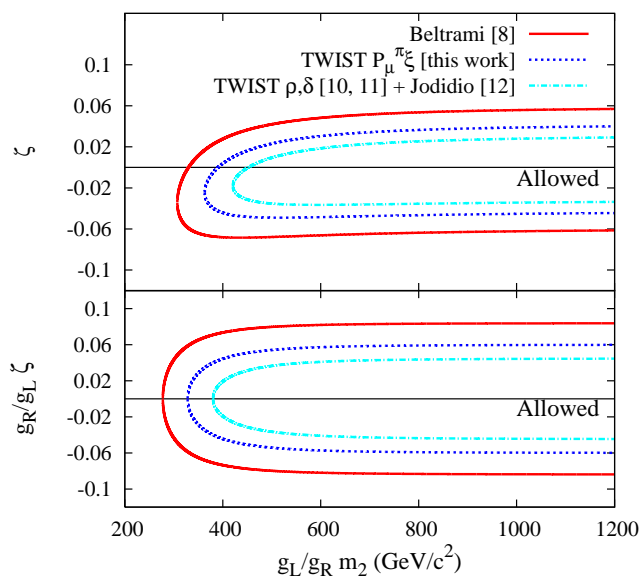}
  \caption[Limits on the manifest LRS model from measured
    $P_{\mu}^{\pi}\xi$]{ (colour online) The top panel shows the manifest
    LRS model 90\% confidence limits on $\zeta$ and $m_2$
    ($g_L/g_R=1$) from measurements of $P_{\mu}^{\pi}\xi$.  The bottom panel
    shows the same limits in the general LRS model case. }
  \label{fig:lrsfinal}
\end{figure}

The 90\% confidence limits on the LRS model parameters from this
measurement are: $-0.050<\zeta<0.041$ and $m_2>360$ GeV/c$^2$ in the
manifest case, and $-0.061<g_R/g_L \zeta<0.061$ and $g_L/g_R m_2>325$
GeV/c$^2$ in the general LRS model.  The LRS model limits are shown in
Fig. \ref{fig:lrsfinal}.

The central value measured is closer to the SM value than previous
direct measurements, and, hence, in a global fit with all other muon
decay parameter data \cite{Gagliardi} it pulls those parameters that
are sensitive to $P_{\mu}^{\pi}\xi$ ($Q_{RR}$, $Q_{LR}$) closer to the
SM value.  However, the changes are small compared to the uncertainty
on these parameters.  The present result reduces the uncertainty on
the direct measurement of $P_{\mu}^{\pi}\xi$ \cite{Beltrami} by a
factor of two; it is also consistent with the SM and the value
obtained indirectly \cite{Jodidio,Musser,Gaponenko}.  This is TWIST's
first independent measurement of $P_{\mu}^{\pi}\xi$, and prospects for
reducing the main systematic uncertainties in $P_{\mu}^{\pi}\xi$ for
data taken in the future with an improved beam measurement device are
excellent.

\begin{acknowledgments}
We would like to thank C.A.~Ballard, M.J.~Barnes, J.~Bueno, S.~Chan,
B.~Evans, M.~Goyette, A.~Hillairet, K.W.~Hoyle, D.~Maas, J.~Schaapman,
J.~Soukup, C.~Stevens, G.~Stinson, H.-C.~Walter, and the many
undergraduate students who contributed to the construction and
operation of TWIST.  We thank D.G.~Fleming and J.H.~Brewer for
discussions on muon spin interactions in condensed matter.  We also
acknowledge many contributions by other professional and technical
staff members from TRIUMF and collaborating institutions.  This work
was supported in part by the Natural Sciences and Engineering Research
Council and the National Research Council of Canada, the Russian
Ministry of Science, and the U.S. Department of Energy.  Computing
resources for the analysis were provided by WestGrid.
\end{acknowledgments}

\end{document}